
\magnification = \magstep1
\baselineskip = 24 true pt
\hsize = 16.5 true cm
\vsize = 23 true cm

\rightline { ICTP Preprint, 1993 }
\vskip 0 true cm
\rightline { IC/93/218 }
\vskip .5 true cm
\centerline { \bf Quantum group and Manin plane }
\centerline {\bf  related to a coloured braid group representation }
                   \vskip 1.5  true cm
\centerline  { B. Basu-Mallick$^{*{\dagger  }  }$   }
\centerline { International Center for Theoretical Physics }
\centerline { I.C.T.P. , P.O. Box 586, 34100 Trieste , Italy  }
                 \vskip 2.25 true cm

\noindent { \bf Abstract }

By considering   `coloured' braid group  representation
we have obtained a   quantum group, which reduces to
the standard $GL_q(2)$ and $GL_{p,q}(2)$ cases  at  some
particular limits of the `colour' parameters.
In spite of   quite
 complicated nature, all of these new quantum group relations
 can be expressed neatly in the Heisenberg-Weyl form, for
 a nontrivial  choice of  the   basis elements.
 Furthermore,  it is  possible to  associate invariant Manin planes,
parametrised by the `colour' variables,  with such quantum group
structure.
\vskip 3.5 true cm
\hrule
\noindent ${}^{* }$Permanent address:
T.N.P.  Division,
  Saha Institute of Nuclear Physics,
 Block AF, Sector 1, Bidhannagar, Calcutta-700 064, India . \hfil \break
\noindent  e-mail address: biru@saha.ernet.in   \hfil \break
\noindent ${}^{\dagger } $Address after November, 1993 :
The Institute of Mathematical Sciences, CIT Campus, Tharamani,
Madras 600113, India.
\hfil \break
\vfil \eject

\noindent { \bf 1. Introduction }
\vskip .2 true cm
In recent years quantum groups and related algebras are found to have a wide
range of applications in different
 branches of physics and mathematics [1-12].
In particular, these algebraic structures manifested themselves in the study
of quantum  integrable models, as an abstraction   of some basic
relations like quantum Yang-Baxter equation [3,10-11].
  From the mathematical point of view, quantum groups are also intimately
  connected to the braid group representations. To illustrate this,
   one may recall the case of well known $GL_q(2)$ quantum group
 generated by the elements
  $~a,~b,~c,~d, ~$ which  satisfy   the  algebraic relations
  $$
     \eqalign {
     ab~=~q^{-1}  ba~,~~ac~=~q^{-1}  ca~,
     ~~&db ~=~ q  bd ~,~~dc ~=~q  cd~, \cr
          bc ~=~ cb~,~~~[a,d] ~=&~ - ~( q-q^{-1} )  cb ~,   } \eqno (1.1)
 $$
  where the deformation parameter $q$ is a nonzero complex number.
  Remarkably,  the above bilinear relations can be expressed
  in a compact matrix form as [3]
  $$           R~T_1~T_2 ~~=~~ T_2~T_1~R    ~,   \eqno (1.2)        $$
where $T$ is a $(2 \times 2 ) $-matrix given by
  $$  T = \pmatrix { {a} & {b} \cr {c} & {d}  } ,     \eqno (1.3) $$
$T_1 = T \otimes {\bf 1} , ~ T_2 =  {\bf 1} \otimes T ~$  and $R$ is a
$(4 \times 4 ) $-matrix  with usual $c$-number matrix elements:
$$
R ~=~ \pmatrix {
{ q }     & {} & {} & {} \cr
{  } &  { 1 }  &  { (q-q^{-1} ) }  & {} \cr
{} & {0} & { 1 }  &  {} \cr
{} & {} & {} & { q }    } ~.
\eqno (1.4)
$$
The expression  (1.2) reveals that
 $~ \Delta T = T { \buildrel  . \over \otimes  }T ~$
would be a  coproduct of  $GL_q(2)$
quantum group, where the symbol  ${ \buildrel  . \over \otimes }$
signifies ordinary matrix multiplication  with tensor multiplication
of algebra.

As it is well known, the form of the algebraic relation (1.2) is quite
general and can be applied to generate other quantum groups also,
depending on the choice of
 the corresponding $R$-matrix. Due to  associativity
 of algebra (1.2),  the $R$-matrix   in general  satisfy
 the   spectral parameterless   Yang-Baxter equation (YBE)
given by
 $$    R_{12} ~R_{13} ~R_{23} ~~=~~R_{23} ~R_{13} ~R_{12} ~,
\eqno (1.5) $$
where we have used the standard direct product notation ( like
$R_{12} = R \otimes {\bf 1} $ etc.).
  YBE (1.5), in turn, leads to a braid group
representation (BGR) for  the matrix
 $~{\hat R}^+ = {\cal P}R^+ ~$  ( $ {\cal P} $ being
the permutation operator with the property
$~{\cal P} A\otimes B = B \otimes A   {\cal P}  $ ) :
$$
{\hat R}_{12} ~{\hat R}_{23} ~{\hat R}_{12} ~~=~~ {\hat R}_{23} ~
     {\hat R}_{12} ~   {\hat R}_{23} ~.
$$
We   shall call  however
 the $R$-matrix itself as a BGR in what follows,
  for the sake of convenience and to emphasise the close connection
  between them. Thus
 from the above discussion  one finds that BGRs  play a rather
 important role in constructing quantum groups through the
 defining relation  (1.2). In this context  we may mention
about  another $(4 \times 4 )$ $R$-matrix, which
 contains two arbitrary parameters $p,~q$ and
 reduces to (1.4) at the limit $~ p=q $ [13]. This BGR eventually
leads to a $p,q$-deformed $GL_{p,q}(2) $ quantum group, which
has been studied extensively   from  different  viewpoints [7,14-16].

Quite recently   some `coloured' generalisations of BGR
  have interestingly   appeared in the
 literature [17-20], which satisfy
           $$
R_{12}^{ (\lambda , \mu ) }~  R_{13}^{ (\lambda ,\gamma ) }~
R_{23}^{ (\mu ,\gamma ) }~ =~
                            R_{23}^{ (\mu ,\gamma ) }
     ~R_{13}^{ (\lambda ,\gamma ) } ~R_{12}^{ (\lambda , \mu ) } ~,
\eqno (1.6)
 $$
where $\lambda ,~ \mu ,~\gamma ~$ are continuously variable  `colour'
   parameters. Though usually
$~{\hat R}^{(\lambda ,\mu )} = {\cal P} R^{(\lambda ,\mu )} $
is defined as the coloured BGR (CBGR), we would call
 $R^{+(\lambda ,\mu )}$-matrix itself as CBGR
 in analogy with the   previous standard case.
   Now it is natural to enquire whether these CBGRs
   would also lead to a new class of quantum group relations.
   Furthermore, similar to the case of usual quantum groups,
    it should be much encouraging to investigate various
   mathematical and physical properties
   associated to  such algebraic  structures.
     In the present article we like to shed some light
     on these issues by concentrating on a  $(4 \times 4 )$
     CBGR given by
     $$
R^{ (\lambda , \mu ) } ~=~ \pmatrix {
{ q^{1- (\lambda - \mu ) }    } & {} & {} & {} \cr
\
{ \eqalign { \cr  } }
&  { q^{ \lambda + \mu  } }  &  {  (q-q^{-1} ) s^{- (\lambda - \mu ) }   }
& {} \cr
{} & {0} & {  q^{- ( \lambda + \mu ) }  }  &  {} \cr
{} & {} & {} & { q^{1 + (\lambda - \mu ) } }    } ~,
\eqno (1.7)
$$
which might be obtained from the fundamental representation of
universal $R$-matrix related to $U_{q,s}(gl(2))$ quantum algebra  [19].
It is  intriguing to notice that at the limit
$ \lambda = \mu = 0 ,$
the above CBGR  reduces to the  BGR (1.4) associated to $GL_q(2)$
quantum group.  On the other hand, for
$ \lambda = \mu \neq 0 $ we will recover the two parameter dependent
BGR [7,13] corresponding to  $GL_{p,q}(2)$ quantum group. So the
 quantum group which we are  hoping  to obtain  at present
  through the  CBGR (1.7),
should  be  some  `coloured'
 generalisation of both $GL_q(2)$ and $GL_{p,q}(2)$  case.
In sec.2 of this paper we shall first review the approach of ref.19
for generating  the CBGR (1.7) and after that discuss how such CBGR,
as well as its generalisations, might be obtained from the standard
BGRs by `colouring' them through some symmetry transformation of
YBE.

Subsequently in sec.3 we attemt to construct the quantum
group related to the CBGR (1.7). Since now colour parameters are present
 in the CBGR, the defining relation of the
 standard  quantum group (1.2) should
also be modified in a consistent way.
Fortunately, such modified version
already exist in the literature
and  was used to  explore quantum groups related to infinite
dimensional $Z$-graded vector spaces [21]. For the present purpose,
 we also
fruitfully use this modified version and  write down  explicitly
the quantum group relations corresponding to the CBGR (1.7).

Next we turn our attention to some interseting features possessed by
the usual quantum groups and investigate whether these features
remain meaningful   even for the coloured case.
For example, it is worth observing that the set of all
algebraic relations corresponding to both $GL_q(2)$ and $GL_{p,q}(2)$
 quantum groups can be recast in the Heisenberg-Weyl form, for
 unimodular values of the deforming parameters $q$ and $p$ [22,16].
Surprisingly  we also find in sec.3 that,
 in spite of their much complicated
nature,  all independent bilinear relations
appearing in the quantum group related to the CBGR (1.7) can be expressed
finally in the Heisenberg-Weyl form. This fact might be useful
in building up  representations for this coloured case.

Another salient feature of the  standard quantum groups
is their close connection to noncommutative geometry [5-7, 23].
Quantum group structure emerges in fact in a natural way if one consider
transformations on the noncommutative vector space or Manin plane,
which preserve the form of  algebra of the co-ordinates.
So the stimulating  question arises whether there exist some `coloured'
Manin planes  on which the quantum group related to the CBGR (1.7)
acts as endomorphism, i.e., generates transformations which preserve the
related algebraic structures. In sec.4 we seek answer to this question
and find   the  existance of  such Manin planes.
 Sec. 5 is the concluding section.

\vskip .75 true cm
\noindent {\bf 2. Construction of  CBGR }
\vskip .2 true cm
     As it is well known for a quasitriangular Hopf algebra ${\cal A}$,
there exists an invertible universal ${\cal R}$-matrix ( ${\cal R} \in
{\cal A} \otimes {\cal A}   $ ) such that it interrelates comultiplications
 $\Delta ,~{\Delta}'$ through ${\Delta}(a){\cal R}={\cal R} \Delta' (a), $
 where $a \in {\cal A} $  and satisfies the following conditions
 $$
   ( ~id \otimes \Delta ~) { \cal R } ~=~{\cal R }_{13} {\cal R }_{12}~,~~
( ~\Delta \otimes id ~) {\cal R } ~=~{\cal R }_{13} {\cal R }_{23}~,~~
    (~ S \otimes id ~) {\cal R }  ~=~ {\cal R }^{-1}  ~,
$$
 $S$ being the antipode. The above relations also imply that the $R$-matrix
 would be a solution of  YBE (1.5).

 If one  considers now
 the case of $U_q(gl(2))$ quantised algebra, apart from the usual
 generators $S_3,~S_{\pm }$ of $U_q(sl(2))$,  a central element or
  Casimir like operator
 $\Lambda $ is included in the picture with the commutation relations [7]
$$
  \eqalign {
[~S_3,S_{\pm }~] ~=~ \pm ~ S_{\pm } , ~~
[~S_+,S_-~] ~=~ { \sin (2\alpha S_3)  \over \sin \alpha } \cr
 \
 [\Lambda , S_\pm ] ~=~  [\Lambda , S_3 ] ~=~0 ~;~~~~~~~  q=e^{i\alpha } . }
\eqno (2.1)
$$
 As a result the standard comultiplication is also get modified to
 yield
$$
    \eqalign {
    \Delta (S_+ ) ~~&=~~ S_+ \otimes q^{-S_3 } \cdot (qs)^\Lambda
          ~+~   ({s\over q })^\Lambda  \cdot q^{S_3}\otimes S_+ ~,  \cr
    \Delta (S_- ) ~~&=~~ S_- \otimes q^{-S_3 } \cdot (qs)^{- \Lambda }
          ~+~   ({s\over q })^{- \Lambda }  \cdot  q^{S_3} \otimes S_- ~, \cr
  \Delta (S_3 ) ~~=~~ S_3 &\otimes {\bf 1} + {\bf 1} \otimes   S_3 ~, ~~
 \Delta ( \Lambda ) ~~=~~ \Lambda \otimes {\bf 1} + {\bf 1} \otimes \Lambda ~,
}
  \eqno (2.2)
$$
 where  $s$ is  an arbitrary parameter appearing due to the symmetry
 of the algebra.
  The other Hopf algebraic structures like co-unit,
 antipode can be consistently defined and the universal ${\cal R}$-matrix
 may also be constructed as [19]
$$
    {\cal R} = q^{ 2  ( S_3 \otimes S_3  + S_3 \otimes \Lambda -
    \Lambda \otimes S_3  ) } \cdot  \sum_{m=0}^\infty
    \
    {  ( 1-q^{-2} )^m  \over  [m,q^{-2} ] ! } ~ \left ( q^{S_3}
    (qs)^{- \Lambda } S_+ \right )^m  \otimes
    \left (  q^{-S_3} ( {s\over q } )^{ \Lambda } S_- \right )^m ~,
\eqno (2.3)
$$
 where $[m,q] = (1-q^m)/(1-q) $ and $ [m,q]! = [m,q]\cdot [m-1,q] \cdots
 1 $.

 Denoting now the eigenvalue of the Casimir  like
 operator $\Lambda $ by $\lambda $
 and the corresponding $n$-dimensional irreducible representation
 of algebra (2.1) as
 $\Pi^n_\lambda $, we may obtain the `colour' representation
$(\Pi^n_\lambda   \otimes \Pi^n_\mu  ){\cal R} $,
giving  a finite dimensional
CBGR  $R^{(\lambda ,\mu )}$
satisfying (1.6). In particular, for the simple
 two dimensional representation
$\Pi^2_\lambda $   through  identity operator  and Pauli matrices :
 $ ~ \Lambda = \lambda {\bf 1},~ S_3 = {1\over 2} \sigma_3 , ~S_{\pm } =
  \sigma_{\pm }~,~$
 one gets the CBGR  (1.7).

In another recent developement [24,25] similar CBGR,
   as well as its generalisations in arbitrary dimensions,
were obtained  directly
from the standard BGRs by using a symmetry transformation
of YBE. It has been  shown that if $R$-matrix is a solution of eqn.
(1.5) with the `particle conserving' constraint, i.e. its elements
$R_{ij}^{kl}$ are non-zero only when the `incoming particles' $(i,j)$
are some permutations of the outgoing ones $(k,l)$, then one can construct
the  CBGR  $ R^{(\lambda ,\mu  )} $  satisfying (1.6)
with elements given by
$$  \left [ R^{(\lambda ,\mu  )} \right]_{ij}^{kl}  ~~=~~
  R_{ij}^{kl} ~
{   u_l^{(1)}(\lambda )u_j^{(2)}(\lambda )  \over
   u_i^{(1)}(\mu )u_k^{(2)}(\mu )   } ~.  \eqno (2.4)  $$
Here the indices $i,j,k,l $ run from 1 to $N$ and $u_i^{(1)}(\lambda),~
u_i^{(2)}(\lambda)$ are $2N$ number of arbitrary  colour parameter
dependent functions.
Starting now  from standard BGR $~R^{\pm }~$ related to the fundamental
representation of $U_q(sl(N))$ [26] :
$$
R^{\pm } ~=~ \sum_i  q^{\pm 1 }  ~e_{ii} \otimes e_{ii} ~+
 ~\sum_{i \neq j} ~\phi_{ij} ~\cdot ~e_{ii} \otimes e_{jj} ~\pm~(q-q^{-1})
 \sum \limits_{ \buildrel  ~i<j  \over  (i>j) } ~
 ~e_{ij} \otimes e_{ji}  ~,
\eqno (2.5)
$$
which evidently satisfies  the `particle conserving' restriction  and
using further (2.4)   one derives the corresponding CBGR as
$$
\eqalign {
            R^{ \pm (\lambda , \mu ) }  ~&= ~
 ~\sum_i ~q^{\pm 1 } ~
{ u_i^{(1)}(\lambda ) u_i^{(2)}(\lambda ) \over
u_i^{(1)}(\mu ) u_i^{(2)}(\mu ) } ~e_{ii} \otimes e_{ii} ~+~
 \
 ~\sum_{i \neq j} ~\phi_{ij} ~
{ u_j^{(1)}(\lambda ) u_j^{(2)}(\lambda ) \over
u_i^{(1)}(\mu ) u_i^{(2)}(\mu ) } ~e_{ii} \otimes e_{jj}   \cr
\
&~~~~~~~~~~~~~~~~~~~~~\pm~(q-q^{-1})
 ~\sum \limits_{ \buildrel  ~i<j  \over  (i>j) } ~
{ u_i^{(1)}(\lambda ) u_j^{(2)}(\lambda ) \over
u_i^{(1)}(\mu ) u_j^{(2)}(\mu ) }    ~e_{ij} \otimes e_{ji} ~ ,    }
\eqno (2.6)
$$
where     elements of  the  matrix $e_{ij}$ are given by
$~ (e_{ij})_{kl} = \delta_{ik} \delta_{jl} $   and
$\phi_{ij} $ are arbitrary constants with the condition
$~\phi_{ij}  \cdot \phi_{ji} = 1 $. Now it is interesting to observe that in
the particular case $N=2$ along  with the choice
$$
   \phi_{12} = 1 ~,~~~u_1^{(1)} (\lambda ) = 1,~~
     u_2^{(1)} (\lambda ) =  (qs)^\lambda  ,~~
     u_1^{(2)} (\lambda ) =  q^{-\lambda }  ,~
     ~u_2^{(2)} (\lambda ) =  s^{-\lambda } ,   \eqno (2.7)
$$
the form of CBGR  $R^{+(\lambda , \mu )}$  in     (2.6)
reduces    exactly   to the CBGR (1.7), which was  obtained earlier
   from the universal ${\cal R}$-matrix
related to $U_q(gl(2))$  in its fundamental representation. Thus
  one finds here a rather convenient  method to generate  CBGRs from the
standard BGRs  in the `particle conserving'   case,
by simply `colouring' the BGRs through a symmetry transformation
of YBE. We may   hope that the CBGR (2.6), with arbitrary $N$,
would  be similarly related to the
  fundamental representation of the $U_q(gl(N))$
quantised algebra. Notice  that other  type
of CBGRs can  also be constructed by restricting the deforming
parameter $q$ at  root of unity [17-18,20], in contrast to the present case
where it is arbitrary.

     Interestingly,  one can Yang-Baxterise the CBGR (1.7)
and generate a solution of YBE  depending on two-component spectral
      parameters [25]. Moreover, realsation of Faddeev-Reshetikhin-Takhtajan
      (FRT) algebra  [3,21] corresponding to this CBGR is also possible.
      Yang-Baxterisation of such FRT algebra leads to   `coloured'
      generalisations of various well known  quantum integrable
      models  like lattice sine-Gordon
      model, Ablowitz-Ladik model etc. [25]. However,
     at present  our aim is
      to focus on the  quantum group relations corresponding
       to the CBGR (1.7) and to explore some related mathematical
       properties.
\vskip 1.75 true cm
\noindent {\bf 3. Quantum group related to a CBGR }
\vskip .2 true cm
         As already mentioned earlier, we need to modify the defining
         relation of usual quantum group (1.2)
         for the case dealing with a CBGR. Such modified relation
         was previously employed to construct quantum groups related
         to infinite dimensional $Z$-graded vector spaces and may
         be expressed as [21]
         $$   R^{(\lambda , \mu )} ~T_1(\lambda )~T_2(\mu ) ~~=~~
          T_2(\mu )~T_1(\lambda ) ~R^{(\lambda ,\mu )}    ~.   \eqno (3.1)   $$
        Notice that the operator valued elements of the matrix $T(\lambda )$
        appearing above are  explicitly dependent on the
        colour parameter and the
        coproduct in this case  might be given by
    $~ \Delta T(\lambda )
    = T(\lambda ) { \buildrel  . \over \otimes  }T(\lambda ). ~$
        Taking now $ R^{(\lambda , \mu )} $ as (1.7), the
        $ (2\times 2)$  $T(\lambda )$-matrix in the form
  $$
    T(\lambda ) =
  \pmatrix { {a(\lambda )} & {b(\lambda )} \cr {c(\lambda )}
  & {d(\lambda )}  } ,     \eqno (3.2)
  $$
     and  inserting them  in (3.1), we  get   the following
     relations among the elements of our coloured version of the quantum
     group:
$$
    \eqalignno {
    a(\lambda )b(\mu ) ~=~ q^{-1+2\lambda }~b(\mu ) a(\lambda ) ~,~~
  & a(\lambda )c(\mu ) ~=~ q^{-1-2\lambda }~c(\mu ) a(\lambda ) ~,~~& (3.3a,b)
  \cr
    d(\lambda )b(\mu ) ~=~ q^{1+2\lambda }~b(\mu ) d(\lambda ) ~,~~
&    d(\lambda )c(\mu ) ~=~ q^{1-2\lambda }~c(\mu ) d(\lambda ) ~,~~&(3.3c,d)
   \cr
    b(\lambda )c(\mu ) = q^{-2(\lambda +\mu ) }~
    c(\mu ) b(\lambda ) ,~~[ a(\lambda )&, d(\mu ) ] =
    - (q-q^{-1})q^{- (\lambda +\mu ) }s^{-\lambda +\mu }~c(\lambda ) b(\mu ),
    & {} \cr &~ &(3.3e,f )    }
$$
  which might be considered as a $\lambda ,~\mu$-dependent generalisation
  of (1.1). Moreover, in addition to (3.3), we also get  some extra
  independent relations  which after a little manipulation may be
  expressed as
  $$
  \eqalignno {
  &a(\lambda )b(\mu ) ~=~(qs)^{\lambda - \mu } ~a(\mu ) b(\lambda ) ~,~~
    a(\lambda )c(\mu ) ~=~(qs)^{-\lambda + \mu } ~a(\mu ) c(\lambda ) ~,
    ~~& (3.4a,b)
  \cr
    &d(\lambda )b(\mu ) ~=~ (qs)^{\lambda - \mu }~d(\mu ) b(\lambda ) ~,~~
    d(\lambda )c(\mu ) ~=~(qs)^{-\lambda + \mu } ~d(\mu ) c(\lambda ) ~,
{}~~&(3.4c,d)
   \cr
   & b(\lambda )c(\mu ) ~=~ s^{-2(\lambda - \mu ) }~
    b(\mu ) c(\lambda ) ~,~~ a(\lambda ) d(\mu ) ~=~a(\mu ) d(\lambda ) ~,
    & (3.4e,f )    \cr
 & a(\lambda ) a(\mu ) ~~=~~  a(\mu ) a(\lambda ), ~~~~~
b(\lambda ) b(\mu ) ~=~ q^{2(\lambda - \mu )}  b(\mu ) b(\lambda ) ~,
&(3.4g,h) \cr
    &c(\lambda ) c(\mu ) ~=~ q^{-2(\lambda - \mu )} c(\mu ) c(\lambda ),~~~~~
d(\lambda ) d(\mu ) = d(\mu ) d(\lambda ) ~. & (3.4i,j)   }
$$
   Observe that for the case of usual groups relations like (3.4)
do not occur at all, since they become trivial
 in the   monochromatic limit  $\lambda
   = \mu $.  Thus the relations (3.3) and (3.4)   define together
   our `coloured' version of quantum group and reduce to the well
   known $GL_q(2)$ case (1.1)  when  $\lambda = \mu = 0 $.
   On the other hand,  by taking the limit  $\lambda = \mu \neq 0 $
   one can  similarly  reproduce
    the two parameter deformed $GL_{p,q}(2)$ quantum group [7].
Notice  that many other interesting relations like
   $$  \eqalign {
   a(\lambda )b(\mu )~&=~ q^{-1 +\lambda + \mu }s^{\lambda - \mu }
   ~b(\lambda ) a(\mu ) ,~~d(\lambda )a(\mu ) ~=~ d(\mu ) a(\lambda ) ,  \cr
    & [ a(\lambda ), d(\mu ) ] ~=~
    - (q-q^{-1})q^{\lambda +\mu  }s^{\lambda -\mu }~b(\lambda ) c(\mu ),  }
   $$
   etc. are also derivable from the basic ones (3.3) and (3.4).

After obtaining the quantum group related to the CBGR (1.7),
we intend to study some of its mathematical properties.
For this purpose,  one may first observe  that  the set of all algebraic
relations (1.1) corresponding to $GL_q(2)$ quantum group can be recast
in the Heisenberg-Weyl form, for unimodular values  of $q$ [22]. To
achieve this a quantum determinant is  usually defined as
$$           D ~=~ ad ~- ~q^{-1} bc ~,       $$
which  can be shown to
 commute with all elements $a,~b,~c,~d$ through (1.1).
Now one can choose a new basis of $GL_q(2)$ with elements $D,~b,~c,~d$,
since  by using the above relation of quantum determinant
the element  $a$ can be
expressed as  ( assuming the invertibility of  $d$ )
   $$             a ~=~ (~ D + q^{-1} bc ~)~ d^{-1}~. $$
Evidently, at  this new basis all algebraic relations corresponding to
 $GL_q(2)$ quantum group take the Heisenberg-Weyl form for unimodular values
of $q$. Similar conclusion can be drawn
even for the two parameter deformed
$GL_{p,q}(2) $  quantum group [16]. However for this case  the quantum
determinant has to be defined in a slightly modified way, and it
  no longer  commutes  with all elements of $GL_{p,q}(2) $.
  The fact that these quantum groups
   can be recast in the Heisenberg-Weyl  form, plays a crucial role
   in finding their representations in terms of commuting pairs of
    canonically conjugate operators and matrices [22,16].

At this point  the natural question arises whether  relations (3.3)
and (3.4), corresponding to our coloured version of quantum group,
can also be  expressed similarly  in the Heisenberg-Weyl form.
To  accomplish this,
 we need to choose first generators $O_i(\lambda )$ ( $i \in [1,4] $ )
such that the following relations  are satisfied :
          $$
     O_i(\lambda ) O_j(\mu ) ~=~ P_{ij}(\lambda ,\mu )~O_j(\mu )O_i(\lambda )~,
{}~~~ O_i(\lambda ) O_j(\mu ) ~=~ Q_{ij}(\lambda ,\mu )~O_i(\mu )O_j(\lambda
)~,
   \eqno (3.5a,b)
   $$
where we have not used any summation convention for repeated indices
and $ P_{ij}(\lambda , \mu ),$   $ Q_{ij}(\lambda , \mu )   $
are $c$-number functions of the colour parameters.
Notice that similar to (3.4), the relations (3.5b) would become trivial
in the monochromatic limit $\lambda = \mu $. Interestingly,
the elements of the matrices $P$ and $Q$ occuring in (3.5a,b) can
be related through the symmetry conditions given by
$$  \eqalign {
  P_{ii}(\lambda , \mu ) ~=~ Q_{ii}(\lambda , \mu ) ~, ~~
P_{ji}(\lambda , \mu ) ~=~{1\over P_{ij}(\mu , \lambda ) } ~,~~
Q_{ij}(\lambda , \mu ) ~=~{1\over Q_{ij}(\mu , \lambda ) } ~,~~\cr
Q_{ji}(\lambda , \mu ) ~=~ Q_{ij}(\mu ,\lambda  )  P_{ij}(\lambda , \mu )
                    P_{ji}(\lambda , \mu )~. ~~~~~~~~~~~~~~~~~~ }
 \eqno (3.6)
 $$
 It may also be observed that
 if one writes down expressions  like
 $$ ~
O_i(\lambda ) O_j(\mu ) ~=~ S_{ij}(\lambda ,\mu )~O_j(\lambda )O_i(\mu )~,~
$$
  then the elements  $S_{ij}$ would be completely determined
through  $P$ and $Q$ matrices: $S_{ij}(\lambda , \mu )$   $ ~=~
  Q_{ij}(\lambda ,\mu ) $ $ P_{ij}(\mu , \lambda ) .$  Now, in analogy with
  the case of $GL_q(2) $ and $GL_{p,q} (2)$ quantum groups, we define
  the quantum determinant for our coloured case as
  $$
  D(\lambda ) ~=~ a(\lambda ) d(\lambda ) ~-~q^{-1 + 2 \lambda }~
  b(\lambda ) c(\lambda )  ~.
  \eqno (3.7)
  $$
Subsequently  by using (3.3) and (3.4), it is not difficult to
 arrive at the following algebraic relations
  $$
  \eqalign {
  a(\lambda ) D(\mu ) = D(\mu ) a(\lambda ),~
b(\lambda ) D(\mu ) = q^{-4 \mu }~ D(\mu ) b(\lambda ),~
c(\lambda ) D(\mu ) = q^{4 \mu }~ D(\mu ) c(\lambda ),& ~ \cr
&(3.8a,b,c)  \cr
d(\lambda ) D(\mu ) ~=~ D(\mu ) d(\lambda )~,~~~
D(\lambda ) D(\mu ) ~=~ D(\mu ) D(\lambda )~. ~~~~~~~~~~~~~&(3.8d,e)    }
$$
 We present the derivation of (3.8a) in   Appendix A  as an illustration.
Thus one finds that, relations of the type (3.5a) can be obtained if
 the elements of the basis are chosen as $~D(\lambda ), ~b(\lambda ),~
 c(\lambda ) $ and $~d(\lambda ) $. However for the coloured case we
 need to get also   relations like (3.5b), one of which  would be of the
form
 $$
 D(\lambda )b(\mu ) = f(\lambda , \mu ) ~D(\mu ) b(\lambda )~,
 \eqno (3.9)
 $$
 where $ f(\lambda , \mu )$  is a $c$-number function.
 Notice that if one substitutes the expression of quantum determinant  (3.7)
 to the above equation,
   elements carrying the `colour' $\lambda $ would occur twice
in  each term of the l.h.s.. On the other hand,
 only one such element  of `colour'  $\lambda $
 would be present in each term of the r.h.s. of (3.9).
 So by using  quantum group relations (3.3) and (3.4),  which preserve
 the number  of  elements of any   particular colour in both sides of the
equation,   it seems to be  unlikely
  to get expressions like (3.9)
    for the general case $\lambda \neq \mu $.

      To have a way out from this difficulty,  let us closely examine
   the expression of $a(\lambda )$ which may be obtained from (3.7) as
   $$
      a(\lambda ) ~=~ {\tilde O}(\lambda )  ~+~
      q^{-1 + 2 \lambda }~O(\lambda )~,
      \eqno (3.10)
   $$
      where
$$~{\tilde O}(\lambda )  ~=~ D(\lambda )d^{-1}(\lambda )~,~~
      O(\lambda ) ~=~b(\lambda )c(\lambda )d^{-1}(\lambda )~,  $$
   and  the existence of inverse of the operator
      $d(\lambda )$ has been assumed  for all values of $\lambda $.
      Now the key
      observation is that, in the  expression of $a(\lambda )$  (3.10)
       the operator $D(\lambda )$ is not appearing individually,
       but as a part of the composite operator  $~{\tilde O}(\lambda )$.
    Consequently,
    we  have  the freedom to choose the basis of  present coloured version of
    quantum group as
    $$
    \eqalign {
    &O_1(\lambda ) ~=~ {\tilde O}(\lambda )~ =~
    D(\lambda )d^{-1}(\lambda )~,\cr
    O_2(\lambda ) = & b(\lambda )~,~~O_3(\lambda ) =c(\lambda )~,~~
     O_4(\lambda ) = d(\lambda ) ~. }
     \eqno (3.11)
    $$
Surprisingly, as we would find below,   with the new chioce of
 basis (3.11) all quantum group relations can be
    expressed  nicely  in the
 desired form (3.5). As a first step towards this,  by using eqns.
 (3.3) and (3.8)
 one can easily obtain the expressions
 $$
   O_1(\lambda )b(\mu ) \!= q^{-1 + 2 \lambda }b(\mu ) O_1(\lambda ),~
O_1(\lambda )c(\mu ) \!= q^{-(1 + 2 \lambda ) }c(\mu ) O_1(\lambda ),~
O_1(\lambda )d(\mu ) \!= d(\mu ) O_1(\lambda ). \eqno (3.12)
  $$
  The only  independent
  relations which are now needed to derive are of the form
  $~ O_1(\lambda ) O_j(\mu ) $  $ =~
   Q_{1j}(\lambda ,\mu ) $   $ O_1(\mu )O_j(\lambda )~$.
   At this point we notice that by using eqns. (3.3d) and (3.4d)
   it is possible to arrive at the expression
   $$
    c(\lambda ) d^{-1}(\lambda ) ~~=~~
    \left ( { s\over q } \right )^{\lambda - \mu }~c(\mu )d^{-1}(\mu )~.
   \eqno (3.13)
   $$
   It is curious to observe that in contrast to eqns.  (3.3) and (3.4),
     elements of a particular `colour' are present in unequal numbers
   in the l.h.s. and r.h.s. of the above relation.
    By using now  this important relation
 as well as the original ones (3.3)
   and (3.4), we can  obtain the expressions like
         $$
 \eqalignno {
              O(\lambda ) b(\mu ) ~=~
    \left ( { s\over q } \right )^{\lambda - \mu } O(\mu ) b(\lambda ) ~,~~
            &O(\lambda ) c(\mu ) ~=~ q^{-3( \lambda - \mu )}
     s^{- \lambda + \mu  } ~O(\mu ) c(\lambda ) ~, &(3.14a,b)  \cr
        O(\lambda ) d(\mu ) ~=~ q^{-2( \lambda - \mu )}
  ~O(\mu ) d(\lambda ) ~, ~~
       &O(\lambda ) a(\mu ) ~=~ q^{-2( \lambda - \mu )}
  ~O(\mu ) a(\lambda ) ~,  &(3.14c,d) \cr
       a(\lambda ) O(\mu ) ~=~ q^{2( \lambda - \mu )}
  ~a(\mu ) O(\lambda ) ~, ~~&O(\lambda ) O(\mu ) ~=~ O(\mu ) O(\lambda )~,
                                     &(3.14e,f)    }
         $$
where the operator $O(\lambda )$ is defined as in (3.10).
We present the derivation of (3.14a) in  Appendix B and the other  relations
appearing above can also be derived in a similar  fashion.
Now it is  rather easy to verify the validity of the relations
$$
   \eqalignno {
O_1(\lambda ) b(\mu ) ~=~ (qs)^{\lambda - \mu } ~O_1(\mu ) b(\lambda )~,~~
&O_1(\lambda ) c(\mu ) ~=~ (qs)^{- \lambda + \mu }
{}~O_1(\mu ) c(\lambda )~, & (3.15a,b) \cr
O_1(\lambda ) d(\mu ) ~=~ O_1(\mu )d(\lambda ) ~,~~
&O_1(\lambda ) O_1(\mu ) ~=~O_1(\mu )O_1(\lambda ) ~, & (3.15c,d)  }
$$
by first substituting  to them
 $$O_1(\lambda ) ~=~ a(\lambda ) ~-~ q^{-1 + 2 \lambda }~O(\lambda )~, $$
from (3.10) and
 then using the eqns. (3.4a,b,f,g)  as well as (3.14).

   Notice that  the expressions (3.12), (3.15) and those of (3.3),  (3.4)
   which do not contain the operator $a(\lambda )$  play a crucial roal
  for our  purpose. Because
   by starting from them  and exploiting the symmetry
   conditions (3.6), one can finally generate
  all of the  desired     relations (3.5).  Therefore
    we see that for the nontrivial choice of the basis (3.11), unimodular
   values of  the deforming parameters
   $q,~s$ and real values of the colour parameters $\lambda ,~\mu ,$
   the new `coloured'  quantum group  relations
   can also be cast in the Heisenberg-Weyl form (3.5). Moreover, such
   relations form  a complete set, since by inverting them
   one can recover  all of the original ones (3.3) and (3.4).
    Thus from the above discussions one may conclude  that,
    some properties of our
     `coloured' quantum group   are  very similar to that of
  its standard  counterparts.
     In the next section we try to extend this area of similarity
 further, by  investigating whether
     there exist some invariant Manin planes associated to such
     coloured quantum group.
\vskip .5 true cm
   \noindent {\bf 4. Manin plane related to CBGR }
\vskip .2 true cm
    A rather interesting  approach  towards quantum groups is
   found by deforming the coordinates of a vector space
  to be noncommuting objects, which obey a set of  bilinear
   product  relations  [6-7,23] . Then the quantum group might   be
    identified  as  the   operator  which acts  on such noncommutative
      spaces or Manin planes and preserves the form of the algebra of
      coordinates even after transformation. For example,  in the case
      of $GL_q(2)$ group (1.1) we may take the coordinates of a two dimensional
      vector space  as $x_1~,x_2 $  and  with the help of $T$-matrix   (1.3)
      define  a transformation like
      $$
            \pmatrix {  { {x_1}'} \cr {{ x_2}'}  } ~=~
 \pmatrix { {a} & {b} \cr {c} & {d}  } \pmatrix { {x_1} \cr {x_2}  }.
              \eqno (4.1)
      $$
 Now by  assuming the quantum group
  elements to be commuting with the coordinates  and using the relations
  (1.1),
  one can show that ( for $ q^2 \neq -1 $ ) there exist  only two
types  of bilinear product relations between the coordinates,  which
remain invariant under the action (4.1).  These two types of bilinear
relations give us the well known $q$-plane and its exterior
plane, respectively [23]:
     $$
       x_1 x_2 ~=~ q^{-1} ~x_2 x_1 ~;~~
       \xi_1^2 = \xi_2^2 = 0~,~
       \xi_1 \xi_2 ~=~  - q ~\xi_2 \xi_1~.
       \eqno (4.2a,b)
     $$
   Consequently,   the
   transformed coordinates $x'$ and $\xi'$   would also satisfy the
   above form of  commutation relations like $x$  and $\xi $.
   One can also argue on the  other way round  and
                try to find the relations among  the elements
of $T$-matrix (1.3), which would keep the form of  commutation relations
(4.2a,b) invariant under transformation.
 The $GL_q(2)$ quantum group structure (1.1)
emerges in a natural way from such requirement. The two parameter
deformed $GL_{p,q} (2) $ quantum group can also  be obtained in a
similar fashion [7,13,16], by slightly modifying the commutation relations
of the coordinates and its differentials (4.2a,b).

      Now we turn our attention to the `coloured' quantum group
      relations (3.3), (3.4) and attempt to find  out whether there exist
      any invariant Manin plane  related to this case.  Since  at present
      the transformation  matrix $T(\lambda)$  (3.2)  is a function of
  $\lambda $, the coordinates should also  be naturally
        dependent on such colour parameter.
      So in analogy with (4.1), we  may define the transformation as
        $$
     \pmatrix { { {x_1}'(\lambda ) } \cr {{x_2}'(\lambda ) }  } ~=~
 \pmatrix { {a(\lambda )} & {b(\lambda )}
 \cr {c(\lambda )} & {d(\lambda ) }  }
 \pmatrix { {x_1(\lambda ) } \cr {x_2( \lambda )}  }  .
              \eqno (4.3)
      $$
    Next  we make an ansatz of the bilinear product relations  between
    the coordinates  of  different  colours:
   $$
       \eqalignno {
              x_1(\lambda ) x_1(\mu ) ~=~\alpha (\lambda , \mu ) ~
                 x_1(\mu )  x_1(\lambda )~,~~
            & x_1(\lambda ) x_2(\mu ) ~=~\beta (\lambda , \mu ) ~
                 x_2(\mu )  x_1(\lambda )~,  & (4.4a,b)  \cr
        x_1(\lambda ) x_2(\mu ) ~=~\gamma (\lambda , \mu ) ~
                 x_1(\mu )  x_2(\lambda )~,~~ &
         x_2(\lambda ) x_2(\mu ) ~=~\delta (\lambda , \mu ) ~
                 x_2(\mu )  x_2(\lambda )~, & (4.4c,d)  }
  $$
     where
      $\alpha (\lambda , \mu ), ~\beta (\lambda , \mu ) ,~
  \gamma (\lambda , \mu )  $ and  $\delta (\lambda , \mu ) $  are
  $c$-number functions. Notice that at the limit $\lambda = \mu $
associated  to the standard cases, the relations (4.4a,c,d )
  become trivial and from their   consistency requirement  we should have
  $$   \alpha ( \lambda , \lambda )
  = \gamma  ( \lambda , \lambda ) = \delta ( \lambda , \lambda ) = 1 .  $$
Subsequently  we demand that the transformation
 (4.3) for  the coordinates with
  colour $\lambda $   and a similar transformation with matrix
  $T(\mu )$  for     colour $\mu $,
  will keep the  form of the   relations (4.4) invariant.
   In other words, the
  commutation relation between transformed coordinates  would be
  obtained by just replacing  $x_i(\lambda ) $ and  $x_j( \mu ) $
    in (4.4)  by
   $x_i'(\lambda )$ and $x_j'( \mu ) $  respectively.
     Exploiting such invariance condition,
     assuming
   the matrix elements of $T(\lambda )$ to be commuting with the coodinates
   $x_i(\mu )$ for all values of  $\lambda ,~\mu $ and  using
   the coloured quantum group relations (3.4), (3.5),
   it is possible to  get  after a long but straightforward calculation
    two sets of solutions for  the coefficients
$\alpha (\lambda , \mu ), ~\beta (\lambda , \mu ) ,~
                   \gamma (\lambda , \mu ) ,~ \delta (\lambda , \mu ) $
                   in (4.4).
   Denoting the  coordinates of the
    invariant `coloured' planes corresponding to these two
   sets of solutions by $x_i(\lambda ) $  and $\xi_i(\lambda ),$
    we   write down the  desired commutations relations as
     $$   \eqalign {
             x_1(\lambda ) x_1(\mu ) ~=~q^{\lambda - \mu } ~
                 x_1(\mu )  x_1(\lambda )~,~~
            & x_1(\lambda ) x_2(\mu ) ~=~ q^{-(1+\lambda + \mu )} ~
                 x_2(\mu )  x_1(\lambda )~,    \cr
        x_1(\lambda ) x_2(\mu ) ~=~s^{- \lambda + \mu } ~
                 x_1(\mu )  x_2(\lambda )~,~~ &
         x_2(\lambda ) x_2(\mu ) ~=~q^ { -\lambda + \mu } ~
                 x_2(\mu )  x_2(\lambda )~,   }  \eqno (4.5)
    $$
    and
   $$  \eqalign {
      \xi_1(\lambda ) \xi_1(\mu ) ~= ~&  \xi_2(\lambda ) \xi_2(\mu )
{}~=~0~,~~\cr
       \xi_1(\lambda ) \xi_2(\mu ) ~=~-~ q^{1-\lambda - \mu } ~
                 \xi_2(\mu )  \xi_1(\lambda )~, & ~~
        \xi_1(\lambda ) \xi_2(\mu ) ~=~s^{- \lambda + \mu } ~
                 \xi_1(\mu )  \xi_2(\lambda )~.  }
    \eqno (4.6)
    $$
It is interesting to notice that at the limit $\lambda = \mu = 0 $,
 (4.5) and (4.6) reduce to  the  commutation relations
   (4.2a)  and (4.2b) respectively ,
    related to the  $GL_q(2)$ quantum group.  On the other hand, for
    $\lambda = \mu \neq 0 $ one would similarly  recover
    the  commutation relations  of the $q$-plane and its exterior plane [7],
      corresponding  to the two parameter deformed  $GL_{p,q}(2)$  case.
        Thus we surprisingly  find
  that  invariant Manin planes can  also  be attached to the present
     coloured quantum group relations,
        in anagogy with its standard counterparts.
\vskip .5 true cm
\noindent { \bf 5. Conclusion}
\vskip .2 true cm
   In this paper, we have investigated
    the quantum group related to a `coloured' braid
   group  representation (CBGR). Interestingly,
   the well known $GL_q(2)$ and $GL_{p,q}(2)$
  quantum groups can  be   recovered as some special cases, from
 this `coloured' quantum group (CQG).
  In spite of their  quite
 complicated nature, all of these new quantum group relations
 can be expressed neatly in the Heisenberg-Weyl form,  in analogy
 with its standard counterparts.
 However to achieve this,
 a nontrivial  choice of  the corresponding    basis elements
 seems to play a crucial role.
 Furthermore,  it is possible to  associate invariant Manin planes,
parametrised by the colour variables,  with our  CQG
structure.

       These results might have implications in several directions.
Since the  new CQG relations
can be recast in the
Heisenberg-Weyl form, we   hope that  in analogy with the standard cases
its elements  may also  be realised through
mutually commuting   pairs of canonically conjugate operators and matrices.
Such realisations  could be important in the context of quantum
integrable models, if one interprets the `colour parameters'
as the `spectral parameters'.  Moreover, it should   be intersesting
  to study the  CQG  relations corresponding to
  other kind of CBGRs and examine  whether they
can also be   expressed in the Heisenberg-Weyl form.  The possibility
of attaching invariant Manin  planes to the CQG relations might
also be much promising. Such approach may lead us  to a whole
class of noncommuatative  quantum planes parameterised by the colour
variables. However the geometrical interpretation of such
colour parameters seems to be yet lacking and whether one can
build up differential geometry on these `coloured' quantum planes
might be an interesting problem for future study.
\vskip .5 true cm
\noindent {\bf Acknowledgments }

\noindent The author likes to thank Prof. H. Saleur  for useful
discussions. \hfil \break
 He also likes to thank Prof. A. Salam, IAEA and UNESCO for
hospitality at International Centre for Theoretical Physics,
Trieste.
\hfil \break
\vfil \eject
 \noindent {\bf Appendix A. Derivation of the relation (3.8a)  }
 \vskip .25 true cm
          By using first the expression of quantum determinant
          (3.7) we get
  $$  a(\lambda ) D(\mu )
  = a(\lambda ) a(\mu ) d(\mu ) - q^{-1 + 2 \mu } a(\lambda )b(\mu )c(\mu ).
  \eqno (A.1)
  $$
   Now application of the relations (3.4g) and (3.3f) in order  yields
   $$
     \eqalign {
     a(\lambda ) a(\mu ) d(\mu ) &= a(\mu ) a(\lambda )  d(\mu ) \cr
     &= a(\mu ) d(\mu ) a(\lambda )  - (q-q^{-1}) q^{-(\lambda + \mu )}
     s^{-\lambda + \mu }  a(\mu ) c(\lambda ) b(\mu )~. }
                        \eqno (A.2)
    $$
  Next, with the help of the relations (3.3b) and (3.4b) one obtains
  $$ a(\mu ) c(\lambda ) =
   q^{-(1+\lambda +\mu )} s^{ \lambda - \mu  } c(\mu ) a(\lambda ) .
   \eqno (A.3)
   $$
   By using now relations  (A.3), (3.3a) and (3.3e) in order we get
  $$
     \eqalign {
   a(\mu )c(\lambda )b(\mu ) &=
  q^{-(1+\lambda +\mu )} s^{ \lambda - \mu  } c(\mu ) a(\lambda ) b(\mu ) \cr
  &=   q^{-2 + \lambda - \mu }  s^{ \lambda - \mu  }
   c(\mu )  b(\mu ) a(\lambda ) \cr
   &=  q^{-2 + \lambda + 3 \mu }  s^{ \lambda - \mu  }
     b(\mu ) c(\mu ) a(\lambda )  . } \eqno (A.4)
 $$
  Substitution of (A.4) into (A.2) then gives
    $$  a(\lambda )a(\mu ) d(\mu ) = a(\mu ) d(\mu ) a(\lambda ) - (q-q^{-1})
     q^{-2 + 2 \mu }    ~ b(\mu ) c(\mu ) a(\lambda ).
     \eqno (A.5)  $$
On the other hand, by using relations (3.3a) and (3.3b) in order
one obtains
$$
  \eqalign {
a(\lambda )b(\mu )c(\mu ) &= q^{-1+2\lambda }~ b(\mu )a(\lambda ) c(\mu )  \cr
     &= q^{-2} ~ b(\mu ) c(\mu ) a(\lambda )  . }
   \eqno (A.6)
 $$
 Substituting now (A.5) and (A.6)  in the r.h.s. of (A.1)
 we finally arrive at the desired relation (3.8a).
\hfil \break
\vfil \eject
 \noindent {\bf Appendix B. Derivation of the relation (3.14a) }
 \vskip .25 true cm
  The application of the relations (3.3e) and (3.3c)  in order
   yields at first
 $$
 b(\lambda )c(\mu ) d^{-1}(\mu )
 = q^{-2(\lambda + \mu )} c(\mu ) b(\lambda ) d^{-1}(\mu )
 = q^{1 -2\lambda } c(\mu )  d^{-1}(\mu ) b(\lambda ).
 \eqno (B.1)
 $$
 Now by using the definition of operator $O(\lambda )$ from (3.10)
 and applying the relations (B.1), (3.4h), (B.1) and (3.13) one by one
 in order, we would obtain :
 $$
    \eqalign {
   O(\lambda )b(\mu ) &= b(\lambda )c(\lambda ) d^{-1}(\lambda )b(\mu)
   = q^{1-2\lambda }~ c(\lambda ) d^{-1}(\lambda) b(\lambda )b(\mu )  \cr
   &=  q^{1-2\mu } ~c(\lambda ) d^{-1}(\lambda) b(\mu )b(\lambda )
   = b(\mu )c(\lambda ) d^{-1}(\lambda )b(\lambda )  \cr
  &= \left ( {s\over q}  \right )^{\lambda - \mu }
      b(\mu )c(\mu ) d^{-1}(\mu )b(\lambda )  \cr
   & = \left ( {s\over q}  \right )^{\lambda - \mu } O(\mu )b(\lambda )  . }
$$
 \hfil \break
 \vfil \eject

\noindent {\bf References }
\vskip .2 true cm
\item {1.} V.G. Drinfeld, in ICM Proceedings ( Berkeley, 1986) p.798.
\item {2.} M. Jimbo, Lett. Math. Phys. 10 (1985) 63.
\item {3.}   L.D. Faddeev,
N. Yu. Reshetikhin and  L. A. Takhtajan, Algebra
and analysis  I   (1989) 178;
\item {} L.D. Faddeev, in  Fields and Particles,  eds.
H. Mitter et. al. ( Springer-Verlag, Berlin, 1990 )  p.89 .
\item {4.}   M. Wadati, T. Deguchi and  Y. Akutsu,
 Phys. Rep.  180  (1989) 247.
\item {5.} S. Woronowicz, Comm. Math. Phys. 111 (1987) 613.
\item {6.} Y. Manin, {\it Quantum groups and noncommutative
geometry, }  Montreal Univ. preprint, CRM-1561, (1988).
\item {7.} A. Schirrmacher, J.Wess and B. Zumino,
Z. Phys. C49 (1991) 317.
\item {8.} V. Jones, Bull. Amer. Math. Soc. 12 (1985) 103.
\item {9.} E. Witten, Nucl. Phys. B322 (1989) 629; Nucl.
Phys. B330 (1990) 285.
\item { 10.} V. Pasquier and H. Saleur, Nucl. Phys. B330 (1990) 523;
\item {} H. Saleur and J.B. Zuber, {\it Integrable lattice models
and quantum groups, }  Saclay preprint, SPhT/90-071 (1990).
\item {11. } D. Bernard and A. Leclair, Nucl. Phys.B340 (1990) 721.
\item {12.} G. Moore and N. Seiberg, Comm. Math. Phys. 123 (1989) 177;
\item {} L. Alvarez-Gaume , C. Gomez and G. Sierra,
Nucl. Phys. B330 (1990) 347;
\item {} K. Gawedzki, Comm. Math. Phys. 139 (1991) 201.
\item {13.} E.E. Demidov, Y. Manin, E.E. Mukhin and D.V. Zhdanovich,
Prog. Theor. Phys. Suppl. 102 (1990) 203.
\item {14.} A. Sudbery, J. Phys. A23   (1990) L697.
\item {15.} N.Yu. Reshetikhin, Lett. Math. Phys. 20 (1990) 331.
\item {16.} R. Chakrabarti and R. Jagannathan, J. Phys. A24 (1991) 5683.
\item {17.}
Y. Akutsu  and T. Deguchi, Phys. Rev. Lett. 67 (1991) 777 ;
\item {} T. Deguchi  and Y. Akutsu, J.Phys. Soc. Jpn. 60 (1991) 2559.
\item {18.}  M.L. Ge, C.P. Sun and K. Xue, Int. Jour.  Mod. Phys. A7
(1992) 6609.
\item {19.} C. Burdik and P. Hellinger, J.Phys. A25 (1992) L1023.
\item {20.} J. Murakami, in Proc. on Int. Conf. of Euler Mathematical
School, Lelingrad : Quantum groups, 1990 ( Lecture notes in Physics,
Springer Verlag, 1991  )p.350.
\item {21.} L.D. Faddeev, N.Yu Reshetikhin and L.A. Takhtajan,
{\it Quantised Lie groups and Lie algebras ,} LOMI-Preprint,
E-14-87, L:LOMI, 1987.
\item {22.} E.G. Floratos, Phys. Lett. 233B (1989) 395 ;
Phys. Lett. 252B (1990) 97;
\item {} J. Weyers, Phys. Lett. 240B (1990) 396.
\item {23.}  J. Schwenk, in Quantum Groups, eds. T. Curtright, D.
Fairlie and C. Zachos ( World Sciencific, 1991) p. 53.
\item {24.} A. Kundu  and B. Basu-Mallick, J.Phys. A25 (1992) 6307.
\item {25.}  A. Kundu and B. Basu-Mallick,
  {\it Coloured FRT algebra and its Yang-baxterisation
  leading to integrable models with nonadditive $R$-matrix ,}
  Saha  Institute  Preprint  , SINP/TNP/93-05 (1993).
\item {26.} A. Schirrmacher, J. Phys. A24 (1991) L1249.

\end